**Title: Fuzzy Mating Behavior Enhances Species Coexistence and Delays Extinction in Diverse Communities**

**Running head:** Fuzzy mating behavior enhances species coexistence

submitted by

Charles H. Cannon[1,2*] and Manuel Lerdau[1,3]

[1]Key Lab in Tropical Ecology, Xishuangbanna Tropical Botanical Garden, Chinese Academy of Sciences, Menglun 666303, China.

[2]Department of Biology, Texas Tech University, Lubbock, TX 79309 USA. Email: chuck@xtbg.ac.cn

[3]Departments of Environmental Sciences and Biology, University of Virginia, Charlottesville, VA USA. Email: mlerdau@virginia.edu.

[*]Corresponding author: Email: chuck.cannon@ttu.edu. Phone: 1-806-317-7643. fax: 1-806-742-2963.

**ABSTRACT**

Current theories about mechanisms promoting species co-existence in diverse communities assume that species only interact ecologically. Species are treated as discrete evolutionary entities, even though abundant empirical evidence indicates that patterns of gene flow such as selfing and hybridization frequently occur in plant and animal groups. Here, we allow mating behavior to respond to local species composition and abundance in a data-driven meta-community model of species co-existence. While individuals primarily out-cross, they also maintain some small capacity for selfing and hybridization. Mating choice is treated as a 'fuzzy' behavior, determined by an interaction between intrinsic properties affecting mate choice and extrinsic properties of the local community, primarily the density and availability of sympatric inter-fertile species. When mate choice is strongly limited, even low survivorship of selfed offspring (<10%) can prevent extinction of rare species. With increasing mate choice, low hybridization success rates (~20%) maintain community level diversity for substantially longer periods of time. Given the low species densities and high diversity of tropical tree communities, the evolutionary costs of competition among sympatric congeneric species are negligible because direct competition is infrequent. In diverse communities, many species are chronically rare and thus vulnerable to stochastic extinction, which occurs rapidly if species are completely reproductively isolated. By incorporating fuzzy mating behavior into models of species co-existence, a more realistic understanding of the extinction process can be developed. Fuzzy mating strategies, potentially an important mechanism for rare species to escape extinction and gain local adaptations, should be incorporated into forest management strategies.

/body

## Introduction

"The most strange thing about fuzzy species is the way they are ignored." [1]

Numerous ecological mechanisms, both stochastic and deterministic, contribute to species co-existence in highly diverse communities [2–7]. The unified neutral theory of biodiversity and biogeography [8-9] has provided a theoretical null ecological framework to test patterns of species composition and abundance in these communities. In these ecological models, species act as discrete evolutionary units: individuals are inter-changeable with conspecifics but completely distinct from heterospecifics, even if they are inter-fertile. Speciation, in other words, is always assumed to be complete, and individuals of different species are reproductively isolated from one another. Abundant evidence indicates, however, that closely related and sympatric species often maintain some level of inter-specific fertility [10–13]. Numerous studies of secondary contact zones have demonstrated that hybrid offspring can invade new habitat [14], inter-specific gene flow can allow the introgression of advantageous alleles from one species to another [11,15-16], and that genetic rescue through hybridization can occur during population crashes [17]. Self-fertilization, primarily in plants, can also play a significant role in reproduction [18]. Overall, mating behavior appears to be responsive to extrinsic signals and reproductive isolation barriers are probably contextual and fuzzy, instead of being fixed and absolute, either in the individual or population.

While we have achieved a relatively good understanding of ecological mechanisms for species co-existence, we have little understanding of how species diversity evolves and is maintained in tropical communities. These communities differ profoundly from temperate systems in many ways. One unique characteristic of tropical communities is the fact that most species exist sympatrically with several closely-related species. Assuming that the genus level is the taxonomic level at which reproductive isolation is truly fixed, in the Pasoh Long-Term Dynamics Plot [19], only 157 out of 811 species (19%) were the sole representative of their genus in the local community, while another 106 species were sympatric with one other congeneric species. The remaining 548 species (68%) exist sympatrically with at least two other congenerics. This pattern is evident in numerous other locations across Southeast Asia (Fig. 1). Interestingly enough, the most diverse families remain consistent across a large latitudinal range.

In the standard allopatric model of speciation, an elaborate series of biogeographic events where species become subdivided, at least parapatric, and then mix again only after achieving a significant amount of reproductive isolation must be constructed. While sympatric speciation seems possible under certain circumstances, the theoretical requirements are stringent [20-21] and probably not generally found in highly diverse communities, particularly for long lived organisms such as rain-forest trees and coral reefs. In these organisms, the demographic integration of ecological and evolutionary dynamics can extend over centuries [22-24]. For the parapatric model to work in the tropical setting, the selective forces driving ecological speciation would have to be quite strong, because the probability of mixing gene flow with a closely related species is quite high, and the barriers to gene flow would have to evolve more rapidly, during the unusual phases when a species is found in isolation. Examination of species distribution patterns of forest trees across various landscape scales has indicated intermediate to weak associations between species distribution and environmental measures [25-26]. Obviously, a new theoretical model of diversification and diversity maintenance, given the evolutionary and ecological setting in tropical forests, is necessary to explain not only how species co-exist but also how they arose.

**Fuzzy mating behavior**

Here, we use the term 'fuzzy' mating behavior to indicate that an individual's mating behavior is contextual and responds to both intrinsic and extrinsic factors. In our model, the individual's tendency to out-cross with a con-specific, hybridize with an hetero-specific, or self-fertilize depends on both genetic constraints and the composition of the local community. Following the principles of fuzzy logic, we assume that the determinants of mating behavior are the quantity and quality of mate choices (the number of crosses attempted and the relative proportion of self, conspecific, and heterospecific pollen present) and the probability of successful outcomes from those different kinds of crosses. Because tropical trees form the basis for some of the most diverse communities in the world [27-28] and their sessile growth form simplifies spatial analyses of ecological interactions, we focus our model on these organisms, although many aspects of our model apply to most plants and animals. Additionally, fuzzy mating behaviors seem to play an important role for tropical trees, as selfing rates have been shown to be inversely related to population densities [29], particularly in disturbed landscapes [30], while sympatric and closely-related species typically can and do hybridize in natural communities [31-32].

The most diverse genera across Southeast Asia (Fig. 1) exhibit a wide range of reproductive characteristics. The genus *Syzygium* is often the most diverse genus in these forests. Their flowers are large, generally white, open, and showy, while the fruits primarily have dispersal characteristics for bats and birds. Despite being among the most prominent floristic elements, no formal taxonomic treatment exists for the genus [33]. The genus is part of the Myrtaceae family, which also contains the genus *Eucalyptus*, a group well-known for frequent interspecific introgression [34]. Studies indicate that the genomes of eucalypt species are largely co-linear [35], indicating that genomic structure is conserved, which would facilitate fuzzy mating behaviors. Genome size in tropical trees also appears to be largely stable in tropical trees. Polyploidization is generally considered rare, and most genera and even families have consistent ploidy levels, again consistent with the type of flexible behavior proposed here.

Even among genera with clearly divergent floral morphologies, hybridization still occurs [36].

Flowering time has been shown to be slightly staggered among sympatric *Shorea* species [37], but considerable overlap in floral receptivity exists [38]. Self-fertilization, of course, could potentially affect all species, even those with no sympatric close relatives. Although selfing rates are generally poorly known in tropical trees, recent findings suggests selfing may be more frequent than expected [39], particularly for endemic species [40]. For convenience sake, in the following discussion, we refer to hetero-specifics that retain partial inter-fertility as 'near-species'. This idea corresponds to the genic model of speciation [41], where speciation never reaches completion but instead near-species remain partially inter-fertile, experiencing divergent selection on some portion of the genome and low levels of neutral to adaptive gene flow across other parts of the genome.

Highly diverse communities are therefore composed of numerous sets of near-species, where individuals predominantly out-cross but retain some diminished capacity for self-fertilization or inter-specific hybridization. Fuzzy mating behavior in these individuals, instead of being determined solely by genotype, responds to the strong environmental signal contained in the quantity and quality of pollen rain. The quantity of a particular type of pollen, either from conspecifics or near-specifics, could act as a proxy for 'success' in the community of that particular genotype, either because of its capture of the pollination service or its physiological success in that particular location. Pollen precedence would control this behavior, where conspecific pollen is more vigorous [42] and, if present, will fertilize most ovules [43]. On the other hand, if an individual primarily receives hetero-specific pollen, despite reduced vigor and lower fertilization rates, the probability of hybridization correspondingly increases. This type of density-dependent, unidirectional gene flow between sympatric near-species has been reported in oaks [44]. While some plants exhibit some level of self-incompatibility, the selective advantage of self-fertilization typically remains high, despite potential costs of inbreeding depression [45]. In situations where conspecific pollen is rare, possible mentor pollen effects could also facilitate hybridization, as mixtures of self and hetero-specific pollen on the stigma have been found to make the pistil more receptive to hybridization.

Our model incorporates a fuzzy framework of conditions that determine species behavior based on the interaction of several probabilistic realities. While this type of fuzzy reasoning remains somewhat controversial in the scientific literature [46], such conditional logic [47] is widely used by industry and business and have also been usefully applied in other types of ecological modeling of animal behavior [48]. These principles are particularly advantageous when uncertainty plays a significant role in the decision process. Given the general complexity of these highly diverse forest communities, particularly given their slow macro-evolutionary processes [23], we argue that uncertainty plays a central role in the evolutionary and ecological behavior of tropical organisms, particularly in relation to the current selective pressures and the quality and quantity of mates. In these conditions, fuzzy mating behavior, where the strength of the reproductive barriers are contingent upon the environmental signal in the pollen rain, should be an appropriate model.

**Resisting the vortex**

Because a large proportion of species in highly diverse communities are rare [49], most are chronically vulnerable to stochastic extinction [50]. This vulnerability implies either that extinction (and the concomitant immigration and/or speciation) rates are high in such systems or that rare species have coping mechanisms to avoid local extinction. Fuzzy mating behaviors would be a potential coping mechanism because isolated individuals, suffering from an absence of conspecifics and poor pollination service in the local neighborhood, could expand their effective population sizes. Unidirectional gene flow into the rare species could be disadvantageous to individuals of the dominant species if the two species were in direct ecological competition. In highly diverse communities, however, most species form only a small fraction of the entire community. This general feature of highly diverse communities could ameliorate the potential cost of being a genetic donor to a rare species, because direct competition with hybrid offspring clustered around the rare genetic recipient would be correspondingly rare. If direct competition with hybrid offspring is rare, then the genetic costs to the pollen donor are minimal. Previous analysis of density dependent competition at the seedling stage in lowland tropical forest on Barro Colorado Island demonstrated that the individuals were affected by conspecific but not heterospecific competitors [51], but because the majority of species are sympatric with closely-related species, these results underestimate the potential competition among near-species. To explore the potential costs of unidirectional hybridization to common species through direct competition with hybrid offspring of rare species, we examine conspecific and congeneric encounter rates within local neighborhoods in the Pasoh Long-Term Dynamics plot, a lowland rainforest in peninsular Malaysia.

Here, we incorporate fuzzy mating behavior into a meta-community model of species co-existence using a data-based simulation with a simple decision tree based on the quantity and composition of pollen received by an individual. We assume that among near-species, pollination is a stochastic process and the composition of the pollen rain is determined by the relative population densities of inter-fertile individuals in the meta-community. Additionally, we assume that conspecific out-crossing is substantially more likely to produce viable offspring than selfing or hybridization but if a cross is successful, the offspring are ecologically equivalent. This neutrality assumption is central to Hubbell's theory of biogeography and biodiversity. While controversial and often incorrect for ecological time frames, the null model provides a mechanism to examine the outcomes generated by the processes themselves. In our case, over evolutionary time, neutrality is probably a more accurate assumption and even if hybrids and selfed offspring have lowered fitness, the alternative is extinction, which is always more costly. We believe that modeling reproductive output as a fuzzy behavior, controlled by the relativistic local community composition of inter-fertile species and the probability of successful outcomes from various types of crosses, can provide further insight into extinction in highly diverse communities and generate numerous testable hypotheses about the evolution of rare species and has major implications for the management of these species.

## Results

Given stochastic mortality among individuals, we find that hybridization success, self-fertilization success, degree of pollination limitation, and level of fecundity all have significant effects on species co-existence (Fig. 2). Inter-specific hybridization had the most pervasive effect, as even relatively low probabilities of success (~20%), maintained the original community diversity over a substantial number (500x) of reproductive events, regardless of species diversity, pollination limitation, or fecundity. The effect is slightly stronger in more diverse sets of near-species. When pollination limitation is severe (<=10%) and fecundity is high (>10 offspring), self-fertilization strongly promotes species co-existence, even at very low levels of success (10%). When self-fertilization and hybridization are not allowed, the communities quickly lose diversity through stochastic extinction of rare species.

Self-fertilization became the dominant mode of reproduction when pollination limitation is strong (Fig. 3), as almost all of the remaining individuals are a product of selfing at the end of the simulation, even when the success of selfed crosses is only 10%. On the other hand, when pollen limitation is not a factor and individuals can cross with half of the meta-population, selfed individuals never occur in the community (Fig. 3). The degree of pollen limitation does not affect the rate of inter-specific hybridization (the blue line remains the same in the top row of each panel, Fig. 3), due to the fact that the pollen limitation affects the arrival of both conspecific and heterospecifc pollen equally. Finally, the proportion of outcrossed individuals gradually increase with decreasing pollination limitation through the simulation, with the proportion of hybrid offspring reaching a peak and slowly tapering off (Fig. 3). This pattern indicates the while fuzzy mating strategies delays extinction of rare species, they do not prevent extinction.

In terms of possible costs to the common species caused by direct competition with hybrid offspring, we found that closely-related species rarely exist within the same neighborhood in highly diverse forests simply because of the general low densities of species. In the Pasoh forest, the most common tree species accounts for less than 3% of the stems, while species with median levels of abundance contribute less than 0.04% to the entire community. Within a local neighborhood of 25 m radius, a focal individual encounters an average of 197 individuals, composed of an average of 104 species. The observed local neighborhood species diversity is significantly lower than the mean species diversity predicted by a null spatial model (134 spp., $p<0.05$), agreeing with previous reports of clumped species distribution patterns [52]. Overall, the number of stems >1 cm DBH of both conspecifics and congeners represents a very small fraction of the local neighborhood, and focal individuals are more likely to encounter conspecific individuals than congeners (Fig. 4), although as congeneric diversity increases, the encounter rate becomes roughly equivalent for conspecifics and congenerics. Many individuals in genera with 2-3 sympatric species almost never encounter congeneric individuals. Given that individuals almost never directly compete with congenerics, unidirectional inter-specific hybridization, from dominant to rare species, cannot cause significant genetic costs.

## Discussion

Our model demonstrates that allowing individual mating behavior to respond to the composition of the pollen rain arriving from the local community of near-species can delay stochastic extinction of rare species. Individuals of rare species, as their populations crash, shift more of their reproductive effort towards hybridization and selfing because they no longer receive sufficient conspecific pollen. This shift in mating strategy effectively expands the local effective population size of each individual and greatly increases their potential genetic diversification, particularly of potentially adaptive genes present in the dominant species. As the local population approaches the extinction vortex, the fitness costs of fuzzy mating behavior will always be less than the cost of extinction. Additionally, seed dispersal limitation and clumped species distributions are a general characteristic of tropical tree communities [52], which limits the recolonization of a community by rare species and creates spatially clumped "families". The offspring of a rare species, including hybrids, will form a spatially clumped grove. The proximity of these mixed offspring could effectively allow them to interbreed, increasing mate choice with family members and potentially re-establish the local population [17], returning the species to being predominantly out-crossed. During these near extinction events, predominantly out-crossing species could modify their mating behavior to become selfing or hybridizing simply in response to the relative abundance of self, conspecific, and near-specific pollen types that arrive on the plant's stigmas.

These behaviors are probably infrequent, local in spatial scale, and highly episodic, caused by either stochastic or deterministic population decline of one species in the local community of inter-fertile near-species. During these episodes, individuals of an increasingly rare species will primarily receive either hetero-specific or self pollen, greatly increasing the probability and advantage of producing offspring through these alternative reproductive pathways. As would be expected, selfing primarily benefits individuals when pollination limitation is extreme. Even if the average viability of selfed or hybrid offspring is low, a small proportion of individual offspring might have equal or even greater fitness than the mother tree, particularly if environmental factors are changing and creating novel habitats [14]. During these periods of local population decline, inter-specific hybridization can also provide selective advantages to rare species by allowing them to capture advantageous alleles and traits from successful species [16]. If a population is crashing due to deterministic processes, such as susceptibility to fungal infection, then capturing genetic variation and alleles from resistant near-species will be advantageous [53].

Diverse tree communities, exemplified by those in tropical Southeast Asia, exist in highly dynamic biogeographic, climatic, and ecological landscapes [54–56], punctuated by brief periods of rapid change and strong selection that may dominate evolution [57]. High biological diversity itself lends a significant element of ecological complexity to the community, as the local composition of predators, herbivores, pollinators, and other inter-acting species at different trophic levels is variable spatially and temporally. Some studies indicate that known hybrid zones promote biodiversity in other parts of the community [58-59]. Ultimately, given the complexity of these communities, both in their current setting and in their historical dynamics, a high element of uncertainty exists in the reproductive success of an individual genotype, particularly given the general pollination limitation in these highly diverse systems [40], which limits mate choice.

One simple aspect of the model is clear: given stochastic population dynamics and complete reproductive isolation among sympatric species, local extinction of rare species happens fairly rapidly if they are assumed to be reproductively isolated from other near-species. Mitigating processes *must* exist that allow rare species to escape extinction. Little is known about the stability of population size of species in highly diverse communities through long periods of time, but given the longer millenial dynamics, common species in today's forests have probably been rare at times in the past and will certainly be rare in some portion of their range, as tree species are frequently found growing outside their 'preferred' habitat. The benefits of maintaining the capacity for fuzzy mating behavior has probably affected most species in these communities, as no one species truly dominates or probably remains common over millenial periods of time.

Our results have profound implications for the general understanding of species co-existence and the management of rare species in highly diverse communities. Previous work indicates that the processes of competitive exclusion in diverse communities are inefficient and protracted [60]. One fundamental aspect of the neutral theory of biogeography and biodiversity that remains unresolved is the balance between speciation and extinction rates. Recently, a simple insight into the speciation process – that it is not instantaneous - provides realistic estimates of species age and abundance of rare species [61]. The assumption that reproductive isolation between different species ultimately provides a long-term selective advantage to individuals, the basic premise of the Biological Species concept, has been demonstrated in simple scenarios that assume consistent selection pressures in relation to life history strategy, but this basic assumption has *never* been proven for long-lived species in highly diverse communities dominated by high degrees of uncertainty in the selective environment. We argue that while a unique and complex suite of phenotypic traits may provide an 'instantaneous' competitive advantage in a particular ecological community [62], the advantage gained by a particular phenotype over long periods of time is unpredictable and frequently, what was advantageous becomes disadvantageous. Instead, species identity, as determined by its genealogy, may be more fluid and dynamic through time and space, particularly as the rate of change and spatial heterogeneity in the environment increases in relation to the demographic turn-over in the community, particularly if the suite of phenotypic traits are not tightly linked genetically with traits that also promote assortative mating among phenotypes.

Our model applies primarily to species in highly diverse communities, where numerous sets of near-species are embedded within a much larger and diverse community and each population contributes a relatively small proportion to the entire community. Below a certain level of diversity in the community, these mechanisms likely play a minor role, only affecting the small number of species Other major tree genera in Southeast Asian forests include: 1) *Litsea*, in the laurel family, with small, frequently unisexual, and primitive flowers producing largely bird-dispersed fruits; 2) *Aglaia* or *Dysoxylum*, both in the Meliaceae family, which produce profuse displays of minute flowers and produce a wide variety of fruit types; 3) *Ficus*, in the Moraceae, with its highly specialized inflorescence and obligate symbiosis with its pollinating wasps; and 4) *Diospyros*, in the Ebenaceae, in which individuals are mostly unisexual (species are dioecious), flowers are large, open, and showy, and most trees exist in the understory. Several of these groups are also poorly known taxonomically and species identification is quite difficult. Few traits seem to link these diverse plant groups except substantial potential for hybridization in mixed communities, where a large fraction of the species are rare [63], live sympatrically with congeneric species [26-27], and mate choice is frequently limited [64-65]. Most detailed pollination studies in these forests also indicate that generalist pollinators are dominant [66-67]. In less diverse communities, such as temperate forests, successful species truly dominate the community and can gain an advantage through reproductive isolation to prevent the donation of pollen or advantageous alleles to rare species. Examples of persistent inter-specific hybridization in such communities, e.g. oaks, may reflect the influences of other, possibly phylogenetic, factors and may play a smaller role in the maintenance of community diversity [43,68].

Our model of fuzzy mating behavior, based upon relative abundance of conspecifics and near-species, demonstrates that even a small capacity for self-fertilization or hetero-specific hybridization can greatly enhance the ability of rare species to persist in communities where more than two near-species co-exist. The initial number of near-species in the community has relatively little effect on the overall outcome, although richer communities did maintain slightly greater levels of diversity. Given the structure and composition of such communities, the majority of rare species, each interacting with small subsets of inter-fertile congeners could delay local extinction through inter-specific hybridization.

Additionally, our model has implications for the entire community when mate choice is strongly limited because self-fertilization can occur in singleton species without any sympatric relatives. Because no species dominates the community and each species is dispersal limited, little direct competition occurs among near-species, thus minimizing the ecologically competitive costs of introgression and inter-fertility. The episodic nature of these fuzzy mating behaviors could interact simultaneously with other ecological and evolutionary processes and should be incorporated into future models of species co-existence. This evidence strongly suggests that this evolutionary behavior may be an important force in maintaining rare species locally and the management of endangered species should actively encourage gene flow among near-species, to both expand population size and to increase genetic variance in the offspring. Our future is certainly unpredictable given many global trends.

## Material and methods

### Model parameters

We included six variable parameters in our model of fuzzy reproductive behavior and species co-existence. We performed three replicates of all possible combinations of the variable parameters. The variable parameters were:

1. **Species diversity**: the initial diversity of near-species that occur within the larger matrix of the community. The values included 3, 5, and 10. These values capture the majority of the observed range of local diversity of congeneric taxa in the Pasoh 50 ha plot (see below).

2. **Community size**: the numbers of individuals in the community of near-species at the three diversity levels. The relative abundances of each species are based on the patterns observed in Pasoh forest plot data for groups of congeners. 3spp = {195,75,30}; 5spp = {300, 200, 150, 100, 50}; 10spp = {220, 200, 125, 110, 100, 90, 75, 50, 20, 10};
3. **Pollen limitation coefficient (PLC)**: proportion of the community of near-species that act as pollen donors in any one reproductive event. For example, in the three species community simulation where the total population is 300, with a PLC=0.1, a single individual receives pollen from thirty other individuals (0.1 * 300); in the five species community simulation where the total population is 800, with a PLC=0.01, a single individual receives pollen from eight other individuals (0.01 * 800). This parameter had a large impact on reproductive success and was examined across coefficient values ranging from 0.01 to 0.5 {0.01, 0.015, 0.02, 0.025, 0.1, 0.15, 0.25, 0.5}.
4. **Self-fertilization success**: the probability of producing a viable offspring from a self-fertilization event. Conspecific crosses were always successful. A value of 0 signifies complete self incompatibility. These values are derived from empirical studies across tropical tree taxa. {0, 0.1, 0.25, 0.4};
5. **Hybridization success**: the probability of producing a viable offspring from an inter-specific pollination event. A value of 0 signifies complete hybrid incompatibility. These values are derived from empirical studies across tropical tree taxa. {0, 0.1, 0.25, 0.4};
6. **Individual fecundity:** the number of potential progeny that could be produced by an individual in each reproductive event{10, 50, 250}, less the unsuccessful self-fertilization and inter-specific crosses.

Several parameters were not allowed to vary in our simulations. We assumed that all crosses between conspecific individuals were successful and that all viable progeny were equally fit and ecological equivalent, whether they were of self, hybrid, or out-crossed origin. All individuals were bisexual and species identity of the progeny was determined by the mother. The size of the modeled community was fixed through the simulation (see parameter 2). Stochastic mortality was fixed at 1.5%. Note that individual reproductive success and the type of offspring produced can only be determined by the interaction of several probabilistic parameters and community composition and not inherent to the individual.

**Simulation procedure**

We established the initial community for each of three replicates for all unique combination of variable parameters (see above). At the outset of the model, a community of a fixed population size was established (3 spp = 300 individuals; 5 spp = 800 inds., and 10 spp = 1000 inds), composed of 'common' and 'rare' species (see parameter 2). For every individual in the initial community, three values were recorded and tracked through each reproductive event for the entire simulation. These values were: 1) individual identity, 2) species identity, and 3) age.

Each simulation was conducted in an iterative process of three steps (see below). Reproduction was simultaneous for all individuals over 500 events and no spatial effects were incorporated into the model. During each reproductive event, each individual was crossed with a random selection of individuals in the community, irrespective of species, given the values of the pollination limitation coefficient and fecundity parameters. Recruitment was random from the entire pool of viable progeny and equal to the number of stochastic deaths. We did not attempt to model the entire community of nested sets of inter-fertile species but each small subset of sympatric and inter-fertile species was modeled separately. The mean number of species remaining in the community at the end of the simulation from the three replicates for all possible combinations of parameter values was used to generate a response surface to examine the sensitivity of the community at each parameter value across its range of variation.

**Step One**

Generate the progeny for each individual, based upon the particular set of parameter values used in each run of the simulation. Progeny for each individual were generated during each reproductive event in the following way:

A random subset of individuals in the community were chosen as pollen donors, based upon the pollination limitation coefficient (parameter 3).

1. All crosses between conspecifics produce a viable progeny.
2. A subset of inter-specific crosses produce viable progeny, based upon the probabilistic success of inter-specific crosses (parameter 5).
3. The total number of attempted crosses from #1 and #2 are subtracted from the individual fecundity (parameter 6), to obtain the number of remaining ovules for which pollination failed. Self-fertilization was then attempted for this remainder with the number of viable selfed offspring determined by the probabilistic success of self-fertilization (parameter 4).
4. The viable progeny produced by each individual in each reproductive event is therefore the sum of #1, 2, and 3, which will be some fraction of parameter 6.

**Step Two**

A proportion of the existing individuals, chosen randomly, in the model die, given our rate of stochastic mortality (fixed at 1.5%). All viable progeny produced by all individuals in step one are pooled into a 'seed bank'.

**Step Three**

Individuals are chosen from the seed bank as recruits to replace the dead individuals. These progeny are immediately able to reproduce. One 'year' is added to each living individual's age.

**Community composition of closely-related species**

To estimate the potential impact of inter-specific hybridization at the community level, we calculated the number of tree species that live sympatrically in the Pasoh Long-Term Dynamics Plot with congeneric species, assuming that congeneric species are at least partially inter-fertile and represent 'near-species'. To estimate the potential cost of unidirectional gene flow to dominant species, we counted the number of conspecific and congeneric individuals present in the 25 m radius neighbourhood surrounding 601 focal trees, representing fifteen individuals from the ten most common species from four categories of congeneric diversity: 3 species/genus, 3 species/genus, 5-6 species/genus, 10-14 species/genus, plus the most common species in the most diverse genus (45 *Eugenia* spp.). We tested empirical patterns against a null spatial model for local neighborhoods, using 1000 replicates. Simulations and analyses were written in Mathematica 7 [69].

**Acknowledgements**. CHC received funding from the Yunnan Province Science and Technology Talent Project (云南省高端科技人才引进计划 #O9SK051B01) and Xishuangbanna Tropical Botanical Garden of the Chinese Academy of Sciences. MTL received funding from a Senior Foreign Scientist Fellowship from the Chinese Academy of Sciences ( 中国科学院国际合作局 外国专家特聘研究员计划 #2060299). We thank Simon Levin and Ben Blackman for comments on earlier versions. We also thank Cam Webb, Ferry Slik, Sean Rice, and Rob Dorit for helpful conversations.

## Figure Legends

**Figure 1.** Species sympatry and diversity of tropical tree communities in four different locations in tropical Asia. "1 sp." illustrates the number of species with a single representative in the community; "2 spp." illustrates the number with species with at least two sympatric congenerics; and ">2 spp." illustrates the number of species with at least three sympatric congenerics. The five most diverse genera in each location are listed below the solid line, with the total number of species observed indicated.

**Figure 2.** Number of species persisting after 500 reproductive events, given different levels of community diversity, fecundity, pollination success, selfing success and hybridization success (see Methods and Supplemental Information for more details). Each contour line represents 0.5 species and darker shades indicate lower species diversity. Contour lines in each graph are interpolated from selfing and hybridization success rates of 0, 0.1, 0.25, and 0.4.

**Figure 3.** Proportion of community recruitment derived from different types of mating, starting with ten species and large reproductive capacity. A) results of model assuming 0% inter-specific hybridization success; B) 10% success; C) 25% success; and D) 40% success. The green line indicates out-crossed offspring, the red line indicates selfed offspring, and the blue line indicates hybrid offspring. The gray line indicates fraction of remaining species richness, beginning with 10 species. Each small graph represents three replicates for each set of mating parameters through 500 generations. The initial population is completely replaced by new individuals before 200 generations in each simulation. Four values for the success of hybrid crosses are shown in the four larger panels. Four

values for the success of self crosses are shown on a set of four rows in each large panel. Three values for pollination success are shown in each set of three columns in each large panel.

**Figure 4.** The relationship between the number of conspecific and congeneric individuals in a local neighborhood with 25m radius around focal trees. Focal trees represent species with different numbers of sympatric congenerics in the Pasoh 50-ha plot. Each circle represents the mean values for fifteen individuals from 61 of the most abundant species in each suite of sympatric congeners. The diameter of each circle indicates the number of congeneric species present in the community for each species. Several circles are labeled to indicate scale. The solid line illustrates a 1:1 ratio. Mean neighbourhood size was 197 individuals composed of an average of 104 species.

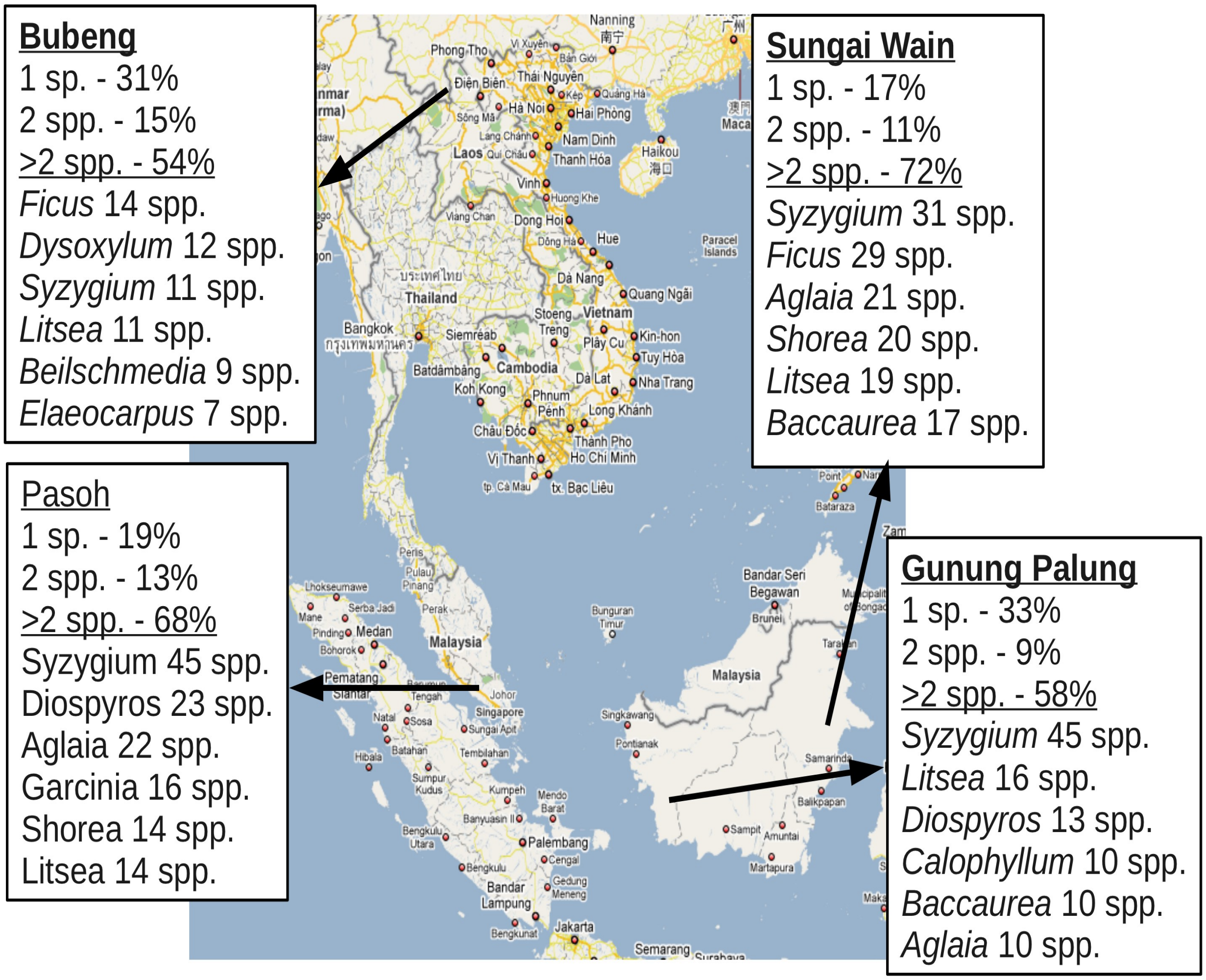
Bubeng
1 sp. - 31%
2 spp. - 15%
>2 spp. - 54%
Ficus 14 spp.
Dysoxylum 12 spp.
Syzygium 11 spp.
Litsea 11 spp.
Beilschmedia 9 spp.
Elaeocarpus 7 spp.
Sungai Wain
1 sp. - 17%
2 spp. - 11%
>2 spp. - 72%
Syzygium 31 spp.
Ficus 29 spp.
Aglaia 21 spp.
Shorea 20 spp.
Litsea 19 spp.
Baccaurea 17 spp.
Pasoh
1 sp. - 19%
2 spp. - 13%
>2 spp. - 68%
Syzygium 45 spp.
Diospyros 23 spp.
Aglaia 22 spp.
Garcinia 16 spp.
Shorea 14 spp.
Litsea 14 spp.
Gunung Palung
1 sp. - 33%
2 spp. - 9%
>2 spp. - 58%
Syzygium 45 spp.
Litsea 16 spp.
Diospyros 13 spp.
Calophyllum 10 spp.
Baccaurea 10 spp.
Aglaia 10 spp.

Figure 1

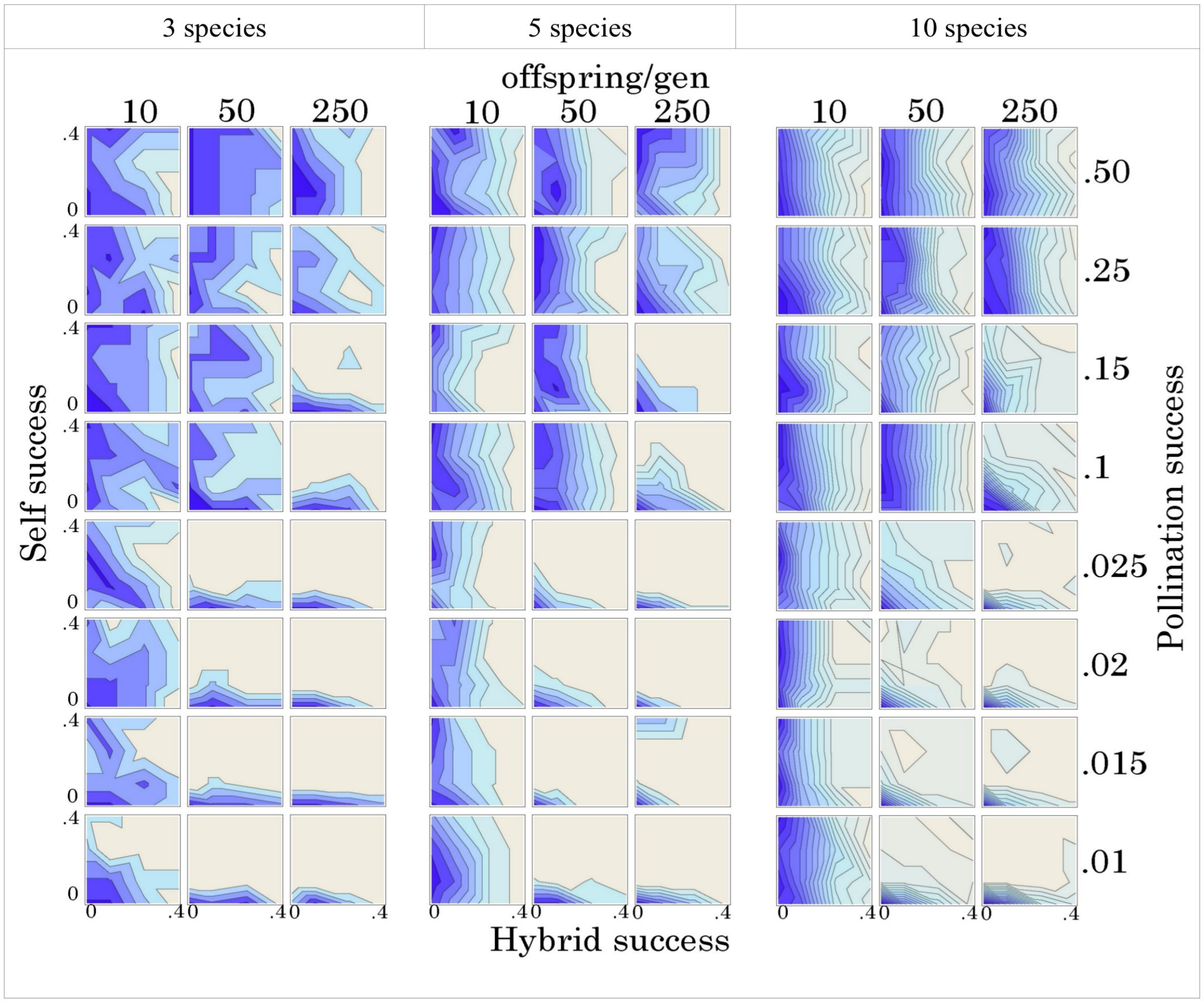

Figure 2

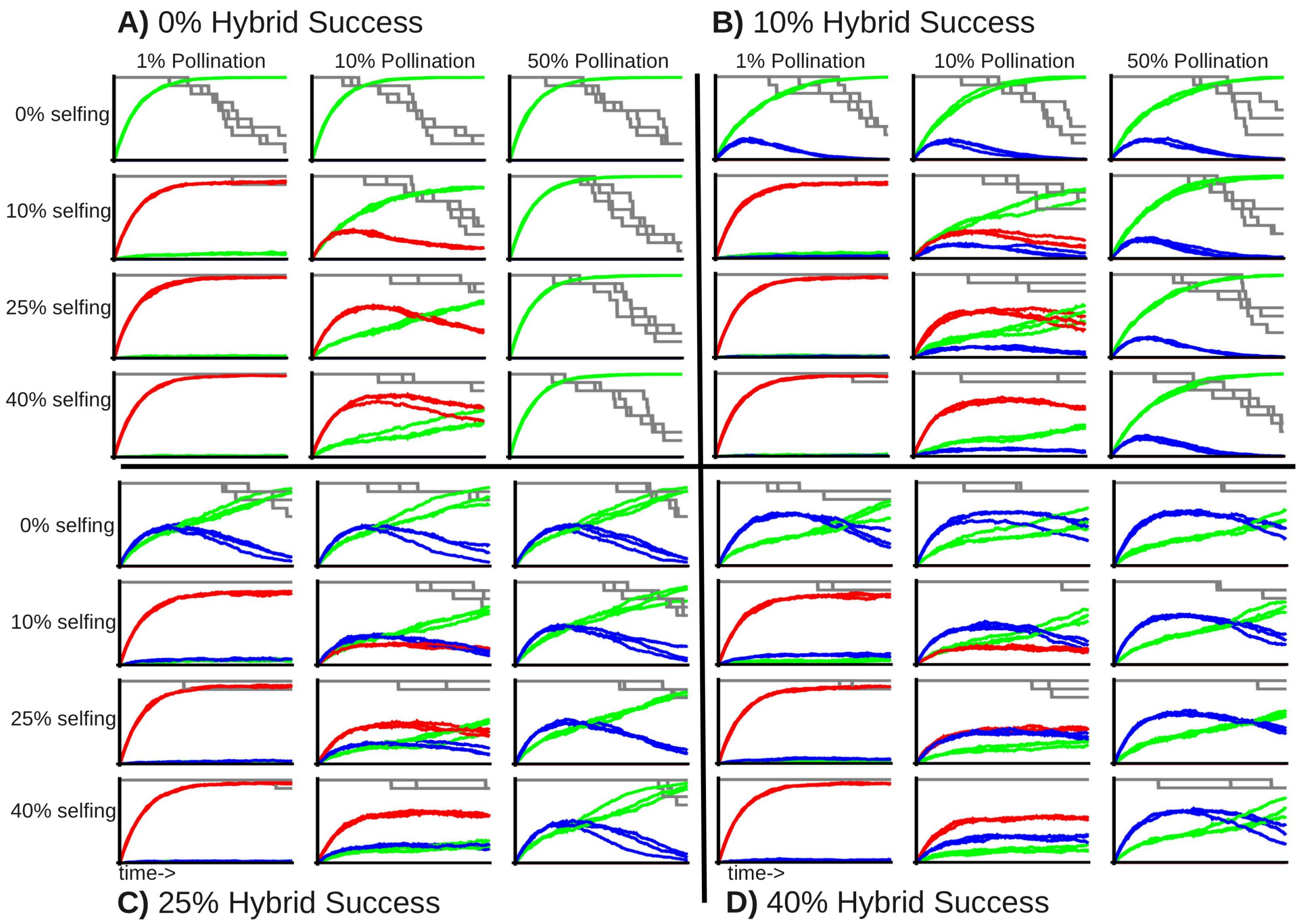

Figure 3

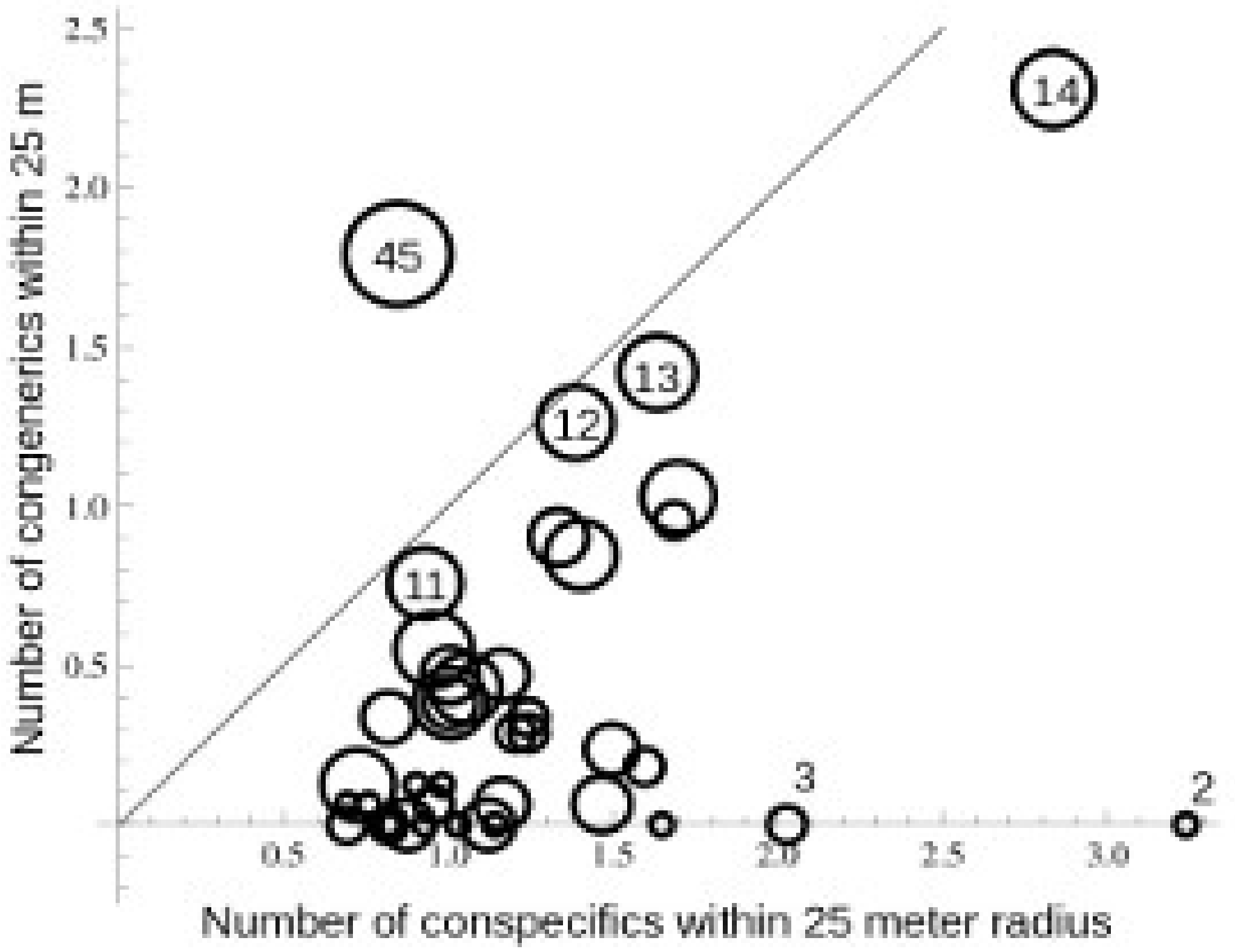

494 Figure 4